\documentclass[twocolumn,twoside,slac_two,amsmath,amssymb]{revtex4}
\pdfoutput=1
\usepackage{graphicx}
\usepackage{fancyhdr}
\usepackage{color}
\begin{document}

\title{The MIDAS Experiment: A New Technique for the Detection of Extensive Air Showers}

\author{C. Williams, A. Berlin, M. Bogdan, M. Bohacova, P. Facal, J. F. Genat, E. Mills, M. Monasor, P. Privitera, L. C. Reyes, 
B. Rouille d'Orfeuil, S. Wayne}
\affiliation{KICP, 5640 South Ellis Avenue, Chicago, IL 60637, USA}

\author{I. Alekotte, X. Bertou}
\affiliation{Centro At\'{o}mico Bariloche and Instituto Balseiro, San Carlos de Bariloche, Argentina}

\author{C. Bonifazi, J. R. T. de Mello Neto, E. M. Santos, }
\affiliation{Universidade Federal do Rio de Janeiro, Instituto de F\'{i}sica, Rio de Janeiro, Brazil}

\author{J. Alvarez-Mu\~{n}iz, W. Carvalho, E. Zas}
\affiliation{Universidad de Santiago de Compostela, Spain}

\begin{abstract}
Recent measurements suggest free electrons created in ultra-high energy cosmic ray extensive air showers (EAS) can interact with neutral air molecules producing Bremsstrahlung radiation in the microwave regime. The microwave radiation produced is expected to scale with the number of free electrons in the shower, which itself is a function of the energy of the primary particle and atmospheric depth. Using these properties a calorimetric measurement of the EAS is possible. This technique is analogous to fluorescence detection with the added benefit of a nearly 100\% duty cycle and practically no atmospheric attenuation. The Microwave Detection of Air Showers (MIDAS) prototype is currently being developed at the University of Chicago. MIDAS consists of a 53 feed receiver operating in the 3.4 to 4.2 GHz band. The camera is deployed on a 4.5 meter parabolic reflector and is instrumented with high speed power detectors and autonomous FPGA trigger electronics. We present the current status of the MIDAS instrument and an outlook for future development.

\end{abstract}

\maketitle

\thispagestyle{fancy}

\section{Introduction}
\label{intro}
The Fluorescence Detection (FD) technique for Extensive Air Showers (EAS) has been well established by the Pierre Auger Observatory (Auger), HiRes, and Telescope Array experiments \cite{augerDetector,hiresDetector,taDetector}.  One key advantage of detecting EAS with FD is the ability to observe the longitudinal development of the shower.  This allows for the measurement of the depth of shower maximum which has proven to be a key parameter in determining the composition of ultra-high energy cosmic rays \cite{augerXmax,hiresXmax}.  

A major drawback to the FD technique is the limited detection duty cycle.  Because fluorescence observations are made in the ultra-violet, the detectors are restricted to only observing on clear, moonless nights.  For example, this represents approximately 13\% of the total observing time for the Auger experiment.\cite{augerDetector}. 

Recent work by Gorham et al. \cite{gorham2008} has shown, through the use of test beam measurements, that it may be possible to observe EAS in the microwave regime using radio detection techniques.  Like the FD technique, detection in microwaves relies on the EAS dissipating energy through ionization.  As the EAS traverses the atmosphere a plasma of free electrons is created.  These free electrons can undergo Bremsstrahlung by interaction with the neutral air molecules producing radio emission in the microwave regime \cite{shower1}.  The radio emission from the EAS plasma is expected to be unpolarized, isotropic, and have a characteristic intensity which scales with the number of particles in the shower.

These characteristics allow for the development of a detection technique that is able to measure the longitudinal development of the EAS.  This would provide a detector which is a direct FD analog with the added benefit of having a nearly 100\% duty cycle.  Also, radio detection in the microwave regime is not affected by atmospheric attenuation or clouds further simplifying the measurement.

\section{Instrumentation}
Recently work has begun at the University of Chicago to build and implement a multi-channel detector on an existing microwave reflector using a combination of commercially available components and custom electronics.  The MIDAS experiment is designed as a stand alone, self-triggered system using a field-programmable gate array (FPGA) trigger and 20MHz analog-to-digital converter (ADC) to find transient events in the microwave regime.

\begin{figure}
\includegraphics[width=65mm]{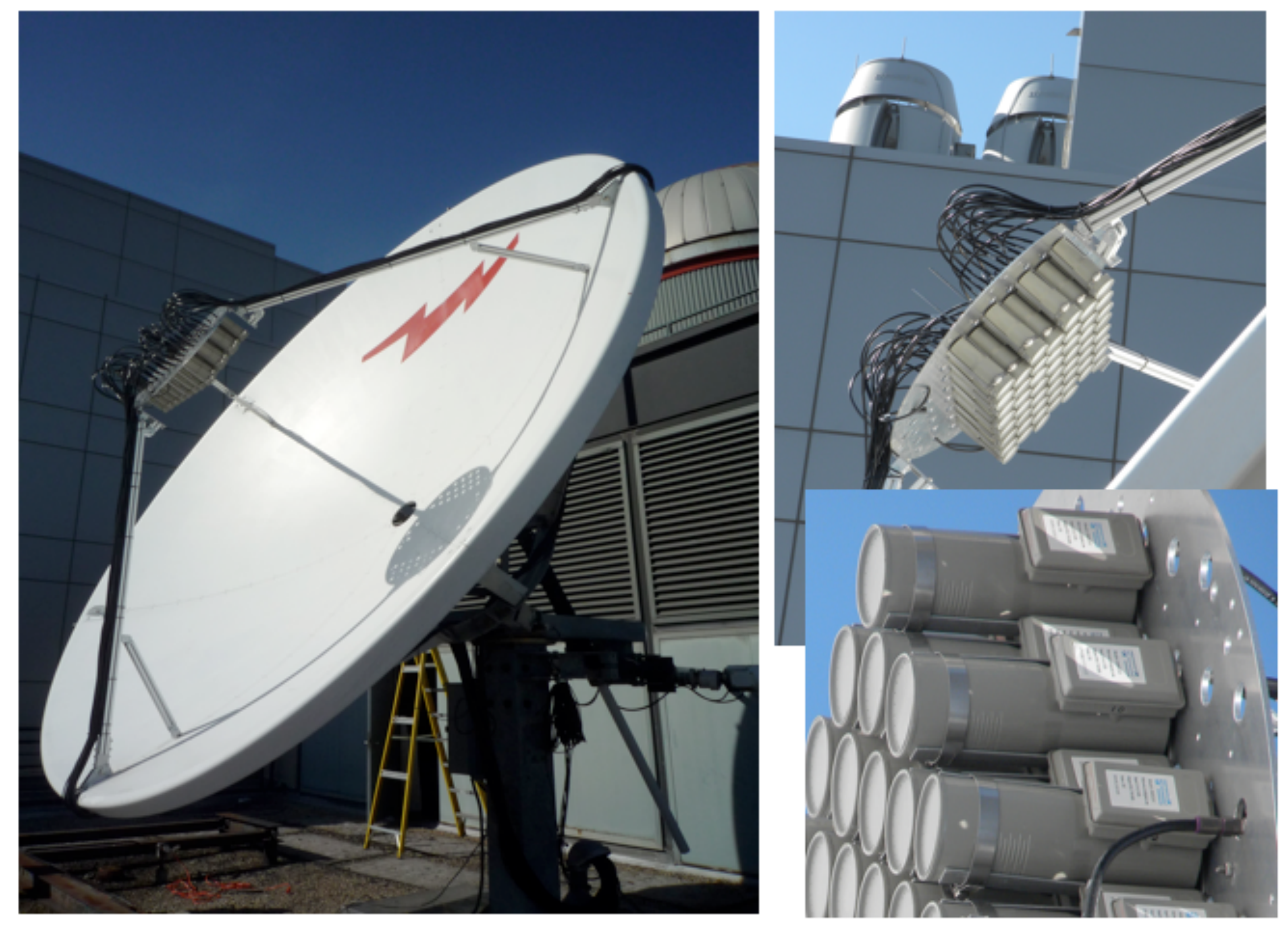}
\caption{Left: 4.5 m parabolic reflector with 53 channel camera mounted at the the prime focus.  Right:  Close-up view of 53 channel camera in reflector's focal plane.}
\label{f1}
\end{figure}

The reflector currently in use, shown in Fig. \ref{f1}, is a 4.5 meter prime focus parabolic dish that is mounted on the roof of one of the Physics Department buildings at the University of Chicago.  The dish is fully steerable with 90 degree range of motion in altitude and 100 degree range of motion in azimuth allowing for pointed observations.  

\subsection{Analog Channel}
The camera currently deployed on the telescope consists of 53 extended C-band feeds operating in the range 3.4GHz to 4.2GHz.  These feeds, shown in Fig. \ref{f1}, are commercially available and designed for use as satellite television receivers.  Each feed is a fully self-contained package consisting of the feed horn, low noise amplifier (LNA), and frequency downconverter.  The feeds can receive in both linear polarizations which are selectable via voltage control.  The LNA operates with a 13K noise floor and a characteristic amplification of 70dB.  The downconverter mixes the input signal down to a range of 950MHz to 1.75GHz allowing for minimal cable loss using standard CATV coaxial cable. 

Signals from each of the 53 feeds are passed through commercially available coaxial power detectors.  The power detectors have a logarithmic power response in the range -55dBm to 0dBm outputting a DC signal between 2V and 0V.  Typical response time for these power detectors is roughly 100ns which is sufficient for EAS detection because a typical shower geometry suggests camera field of view crossing times on the order of tens of microseconds.  

The analog power signals are digitized using custom 14 bit flash ADC boards operating at 20MHz. The electronics boards also contain the FPGA used in the telescope's trigger as described in \S \ref{fpgaTrigger}.  The boards have been developed at the University of Chicago by the Electronics Design Group at the Enrico Fermi Institute.

\subsection{FPGA Trigger}
\label{fpgaTrigger}
\begin{figure}
\includegraphics[width=65mm]{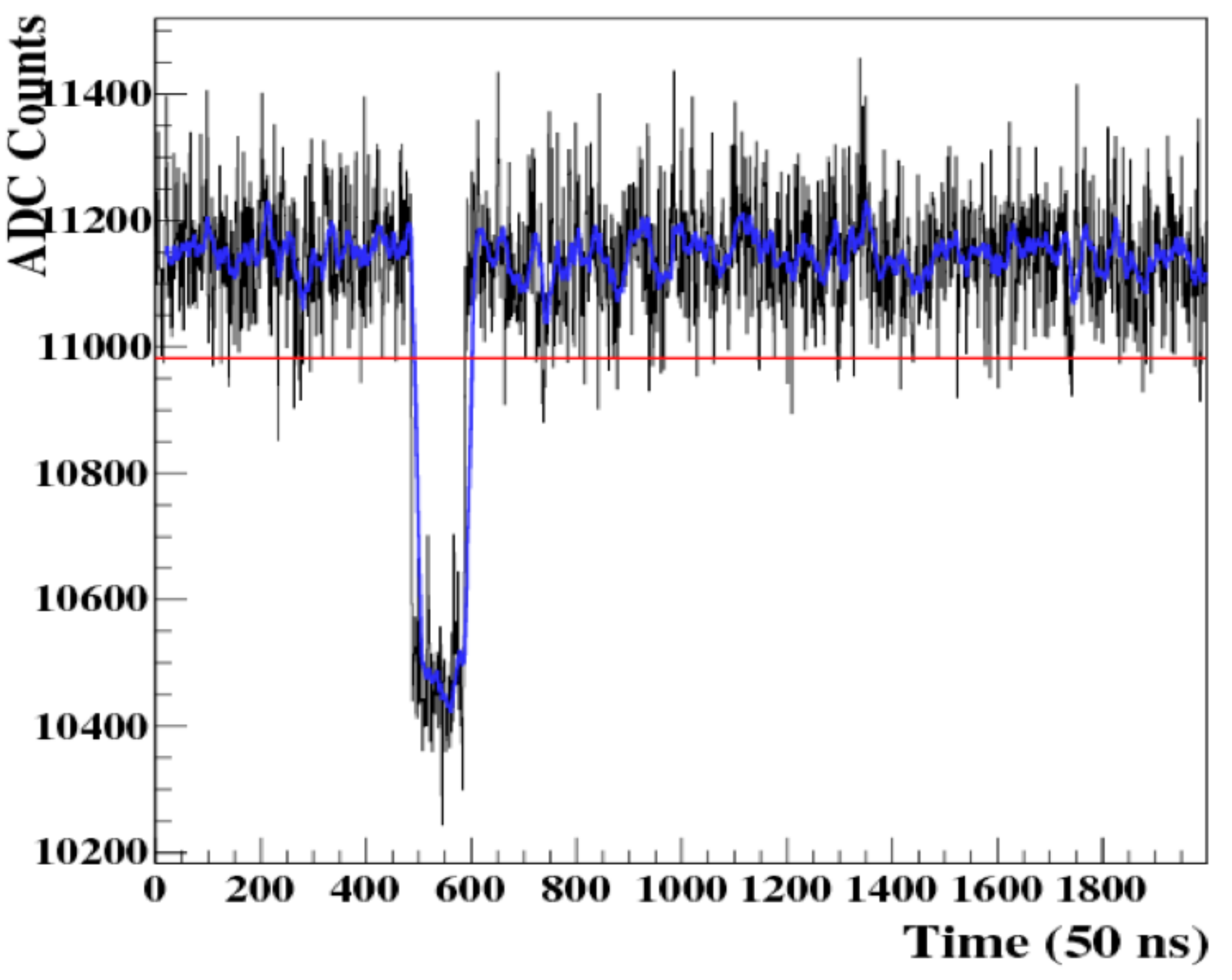}
\caption{Single channel output of a pulse from log-period antenna used for calibration.  Signal is inverted due to reverse voltage bias of power detector.  Black trace is raw output from channel.  Blue trace is the 1$\mu$s running sum.  Red Line represents the threshold level.}
\label{triggered_pulse}
\end{figure}

The FPGA trigger consists of two separate levels of triggering using a 1$\mu$s running sum from each channel.  A first level trigger (FLT) occurs whenever the running sum for a single channel goes over threshold.  Fig. \ref{triggered_pulse} shows an example of a FLT for a single channel which is receiving a pulse from a log-periodic antenna used for calibration.  The threshold for triggering is self-regulated for each channel such that the FLT rate is held fixed at 100Hz.  A second level trigger (SLT) occurs whenever three FLTs forming a specified pattern occur within a time window of 20$\mu$s.  The events are then written to disk.  The three pixel patterns are chosen to reflect possible tracks EAS can make moving through the detector's field of view.  Due to thermal noise fluctuations the random accidental event rate for SLTs is of order 0.1Hz.

The electronics are also implemented with a High Level Veto for periods of time which have a high SLT rate due to anthropogenic noise.  When the SLT rate rises above a preset value the trigger is inhibited and no events are written to disk. 

\section{Calibration}
\label{calib}
By placing a log-periodic antenna at the base of the telescope and directing the beam up towards the camera it is possible to illuminate the feeds nearly uniformly.  Using this antenna to send 5$\mu$s pulses, a relative calibration of the feeds can be made to check both gain values and timing.  Fig. \ref{triggered_pulse} shows an example of one these pulses for a single channel.

The telescope can also be used for pointed observations because it is fully steerable.  A second calibration method that is being used is to observe astrophysical objects with known radio fluxes.  The values observed for these sources can then be checked with the known values.  This provides a way to measure the system temperature.  Currently, observations have been made of the Sun, Moon, and Crab Nebula.  Taking Solar flux in the C band measured by the Nobeyama Radio Observatory, and comparing it with the value measured from the MIDAS system we find a system temperature of approximately 110K.

\section{Detector Monte Carlo}
Alongside the hardware development, a Monte Carlo simulation for the MIDAS detector is also being developed.  The simulation uses a realistic cosmic ray spectrum and geometry.  The molecular Bremsstrahlung emission parameters for simulated EAS are based on values measured by Gorham et al. \cite{gorham2008}.  The beam pattern for each feed and the electronics response are developed from values obtained through calibration measurements outlined in \S \ref{calib}.  

\begin{figure}
\includegraphics[width=65mm]{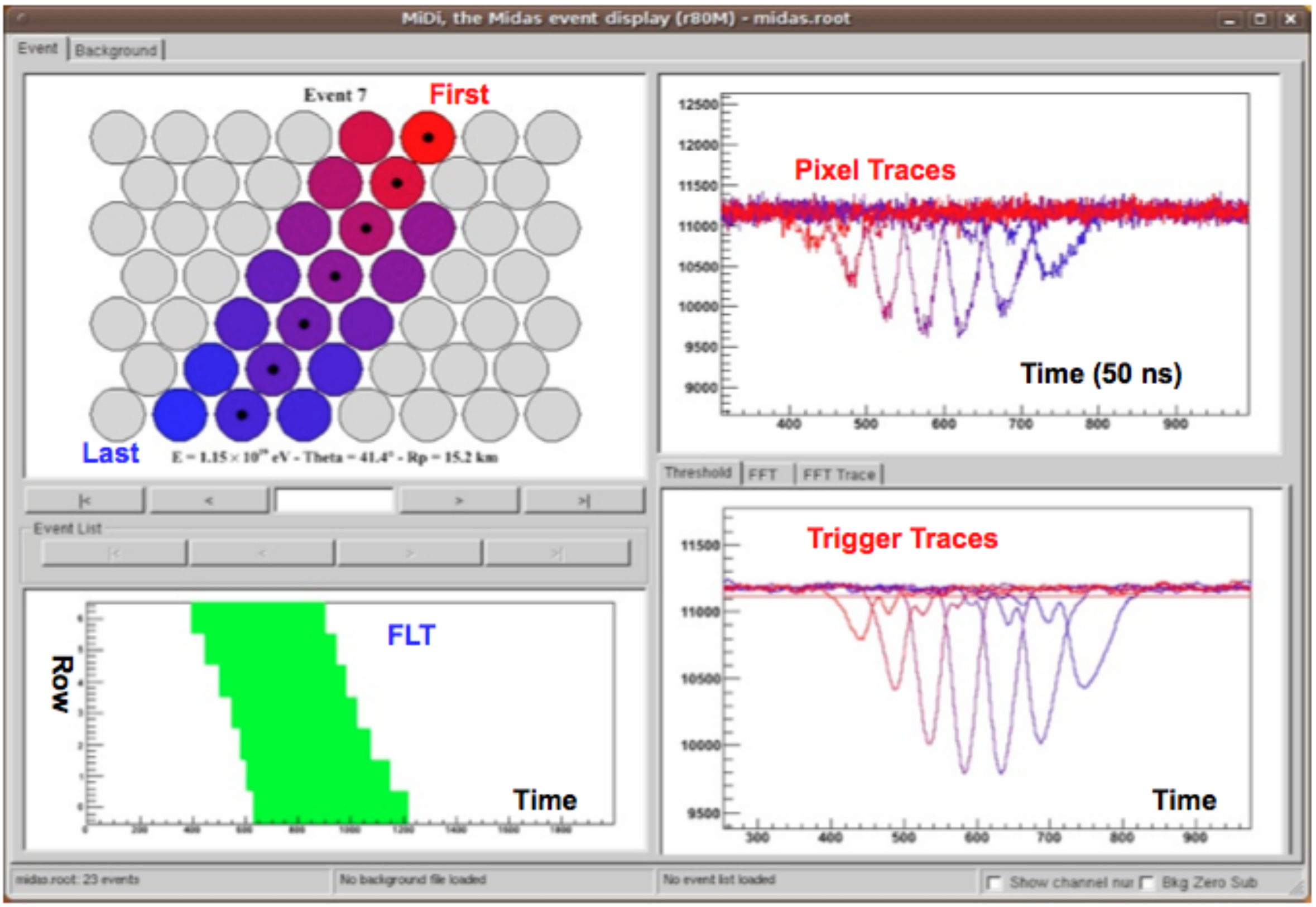}
\caption{Monte Carlo simulated detection of an EAS with the MIDAS experiment.  Shower energy is $1.15\times10^{19}$ eV shower maximum located 15.2km from the detector.  Left panels show time structure and geometry for the event.  The upper right panel displays the raw traces from selected channels and lower right panel displays the 1$\mu$s running sum trace which the system uses for triggering.}
\label{mc_event}
\end{figure}

Fig. \ref{mc_event} shows an example simulation event displayed in the event viewer that has been written for MIDAS.  This Monte Carlo event has an energy of $1.15\times10^{19}$ eV with shower maximum a distance of 15.2km from the detector.  The upper and lower left panels shows the timing structure for the event, the upper right panel displays raw traces, and the lower right panel displays the running sum trace which triggers the detector.

\section{Events}
In the current design the majority of recorded events are thermal noise events that trigger a three pixel pattern.  There is also a large subset of events that we believe are attributed to specular reflection from aviation radio location.  Due the University of Chicago's proximity to Midway International Airport, these events have the potential to produce a high percentage of deadtime in the detector.  A band-stop filter is being developed to help deal with the bursts of noise that lead to a high event rate.

The baseline for each antenna fluctuates due to both the temperature of the LNAs as well as the external noise environment.  The average baseline fluctuation for a period of 40 days of continuous data taking was less than 1dB.  A patch antenna permanently mounted at the base of the reflector, illuminating the feeds will allow baseline values to be systematically monitored over extended time periods.

\section{Conclusion}
We have presented the initial design and commissioning of the MIDAS prototype for detection of molecular Bremsstrahlung emission from EAS in the 3GHz to 4GHz radio band.  The current design of the MIDAS prototype uses a self-triggered multi-feed detector mounted at the prime focus of a parabolic reflector.  Future work will implement improvement to the FPGA trigger logic and the installation of an analog filter to better control for anthropogenic noise, increasing livetime.  Future plans also include installation of the MIDAS system at the Auger site in Argentina to look for events which are coincident with both the FD and surface detectors.

\bigskip
\begin{acknowledgments}
Support for this research at The University of Chicago, Kavli Institute for Cosmological Physics was provided by NSF grant PHY-0551142.
\end{acknowledgments}


\bigskip

\begin{thebibliography}{99} 
\bibitem{augerDetector}
J. Abraham, et al. [Pierre Auger Collaboration], "The fluorescence detector of the Pierre Auger Observatory", Nucl. Instr. Meth. \textbf{A 620}, 227, Aug. 2010.
\bibitem{hiresDetector}
R. U. Abassi, et al. [HiRes Collaboration], "Measurement of the Flux of Ultrahigh Energy Cosmic Rays from Monocular Observations by the High Resolution Fly's Eye Experiment", Phys. Rev. Lett. \textbf{92}, 151101, Apr. 2004.
\bibitem{taDetector}
M. Fukushima. "Measurement of UHECRs by the Telescope Array (TA) experiment", Proceedings of the XVI International Symposium on Very High Energy Cosmic Ray Interactions.  June 2010.
\bibitem{augerXmax}
J. Abraham, et al. [Pierre Auger Collaboration], "Measurement of the Depth of Maximum of Extensive Air Showers above 10$^{\rm{18}}$ eV", Phys. Rev. Lett. \textbf{104}, 091101, Mar. 2010.
\bibitem{hiresXmax}
R. U. Abassi, et al. [HiRes Collaboration], "Indications of Proton-Dominated Cosmic-Ray Composition above 1.6 EeV", Phys. Rev. Lett. \textbf{104}, 161101, Apr. 2010.
\bibitem{gorham2008}
P. W. Gorham, et al., "Observations of microwave continuum emission from air shower plasmas", Phys. Rev. D \textbf{78},  032007, Aug. 2008.
\bibitem{shower1}
G. Bekefi, "Radiation Processes in Plasmas" Wiley, New York,
1966.


\end{thebibliography}
\end{document}